\documentstyle[twoside,epsfig]{article}

\catcode`\@=11
\long\def\@makefntext#1{
\protect\noindent \hbox to 3.2pt {\hskip-.9pt  
$^{{\eightrm\@thefnmark}}$\hfil}#1\hfill}		

\def\@makefnmark{\hbox to 0pt{$^{\@thefnmark}$\hss}}	
	
\def\ps@myheadings{\let\@mkboth\@gobbletwo
\def\@oddhead{\hbox{}
\rightmark\hfil\eightrm\thepage}   
\def\@oddfoot{}\def\@evenhead{\eightrm\thepage\hfil
\leftmark\hbox{}}\def\@evenfoot{}
\def\sectionmark##1{}\def\subsectionmark##1{}}

\oddsidemargin=\evensidemargin
\newcounter{sectionc}\newcounter{subsectionc}\newcounter{subsubsectionc}
\renewcommand{\section}[1] {\vspace{12pt}\addtocounter{sectionc}{1} 
\setcounter{subsectionc}{0}\setcounter{subsubsectionc}{0}\noindent 
	{\tenbf\thesectionc. #1}\par\vspace{5pt}}
\renewcommand{\subsection}[1] {\vspace{12pt}\addtocounter{subsectionc}{1} 
	\setcounter{subsubsectionc}{0}\noindent 
	{\bf\thesectionc.\thesubsectionc. {\kern1pt \bfit #1}}\par\vspace{5pt}}
\renewcommand{\subsubsection}[1] {\vspace{12pt}\addtocounter{subsubsectionc}{1}
	\noindent{\tenrm\thesectionc.\thesubsectionc.\thesubsubsectionc.
	{\kern1pt \tenit #1}}\par\vspace{5pt}}
\newcommand{\nonumsection}[1] {\vspace{12pt}\noindent{\tenbf #1}
	\par\vspace{5pt}}

\newcounter{appendixc}
\newcounter{subappendixc}[appendixc]
\newcounter{subsubappendixc}[subappendixc]
\renewcommand{\thesubappendixc}{\Alph{appendixc}.\arabic{subappendixc}}
\renewcommand{\thesubsubappendixc}
	{\Alph{appendixc}.\arabic{subappendixc}.\arabic{subsubappendixc}}

\renewcommand{\appendix}[1] {\vspace{12pt}
        \refstepcounter{appendixc}
        \setcounter{figure}{0}
        \setcounter{table}{0}
        \setcounter{lemma}{0}
        \setcounter{theorem}{0}
        \setcounter{corollary}{0}
        \setcounter{definition}{0}
        \setcounter{equation}{0}
        \renewcommand{\thefigure}{\Alph{appendixc}.\arabic{figure}}
        \renewcommand{\thetable}{\Alph{appendixc}.\arabic{table}}
        \renewcommand{\theappendixc}{\Alph{appendixc}}
        \renewcommand{\thelemma}{\Alph{appendixc}.\arabic{lemma}}
        \renewcommand{\thetheorem}{\Alph{appendixc}.\arabic{theorem}}
        \renewcommand{\thedefinition}{\Alph{appendixc}.\arabic{definition}}
        \renewcommand{\thecorollary}{\Alph{appendixc}.\arabic{corollary}}
        \renewcommand{\theequation}{\Alph{appendixc}.\arabic{equation}}
        \noindent{\tenbf Appendix \theappendixc #1}\par\vspace{5pt}}
\newcommand{\subappendix}[1] {\vspace{12pt}
        \refstepcounter{subappendixc}
        \noindent{\bf Appendix \thesubappendixc. {\kern1pt \bfit #1}}
	\par\vspace{5pt}}
\newcommand{\subsubappendix}[1] {\vspace{12pt}
        \refstepcounter{subsubappendixc}
        \noindent{\rm Appendix \thesubsubappendixc. {\kern1pt \tenit #1}}
	\par\vspace{5pt}}

\newcommand{\textlineskip}{\baselineskip=13pt}
\newcommand{\smalllineskip}{\baselineskip=10pt}

\def\eightcirc{
\begin{picture}(0,0)
\put(4.4,1.8){\circle{6.5}}
\end{picture}}
\def\eightcopyright{\eightcirc\kern2.7pt\hbox{\eightrm c}} 

\def\abstracts#1#2#3{{
	\centering{\begin{minipage}{4.5in}\baselineskip=10pt\footnotesize
	\parindent=0pt #1\par 
	\parindent=15pt #2\par
	\parindent=15pt #3
	\end{minipage}}\par}} 

\renewenvironment{thebibliography}[1]
	{\frenchspacing
	 \ninerm\baselineskip=11pt
	 \begin{list}{\arabic{enumi}.}
	{\usecounter{enumi}\setlength{\parsep}{0pt}
	 \setlength{\leftmargin 12.7pt}{\rightmargin 0pt} 
	 \setlength{\itemsep}{0pt} \settowidth
	{\labelwidth}{#1.}\sloppy}}{\end{list}}

\newcounter{itemlistc}
\newcounter{romanlistc}
\newcounter{alphlistc}
\newcounter{arabiclistc}

\newcommand{\fcaption}[1]{
        \refstepcounter{figure}
        \setbox\@tempboxa = \hbox{\footnotesize Fig.~\thefigure. #1}
        \ifdim \wd\@tempboxa > 5in
           {\begin{center}
        \parbox{5in}{\footnotesize\smalllineskip Fig.~\thefigure. #1}
            \end{center}}
        \else
             {\begin{center}
             {\footnotesize Fig.~\thefigure. #1}
              \end{center}}
        \fi}

\newcommand{\tcaption}[1]{
        \refstepcounter{table}
        \setbox\@tempboxa = \hbox{\footnotesize Table~\thetable. #1}
        \ifdim \wd\@tempboxa > 5in
           {\begin{center}
        \parbox{5in}{\footnotesize\smalllineskip Table~\thetable. #1}
            \end{center}}
        \else
             {\begin{center}
             {\footnotesize Table~\thetable. #1}
              \end{center}}
        \fi}

\def\@citex[#1]#2{\if@filesw\immediate\write\@auxout
	{\string\citation{#2}}\fi
\def\@citea{}\@cite{\@for\@citeb:=#2\do
	{\@citea\def\@citea{,}\@ifundefined
	{b@\@citeb}{{\bf ?}\@warning
	{Citation `\@citeb' on page \thepage \space undefined}}
	{\csname b@\@citeb\endcsname}}}{#1}}

\newif\if@cghi
\def\cite{\@cghitrue\@ifnextchar [{\@tempswatrue
	\@citex}{\@tempswafalse\@citex[]}}
\def\citelow{\@cghifalse\@ifnextchar [{\@tempswatrue
	\@citex}{\@tempswafalse\@citex[]}}
\def\@cite#1#2{{$\null^{#1}$\if@tempswa\typeout
	{IJCGA warning: optional citation argument 
	ignored: `#2'} \fi}}

\def\pmb#1{\setbox0=\hbox{#1}
	\kern-.025em\copy0\kern-\wd0
	\kern.05em\copy0\kern-\wd0
	\kern-.025em\raise.0433em\box0}

\def\fnt#1#2{\footnotetext{\kern-.3em
	{$^{\mbox{\scriptsize #1}}$}{#2}}}

\def\fpage#1{\begingroup
\voffset=.3in
\thispagestyle{empty}\begin{table}[b]\centerline{\footnotesize #1}
	\end{table}\endgroup}

\font\tenrm=cmr10
\font\tenit=cmti10 
\font\tenbf=cmbx10
\font\bfit=cmbxti10 at 10pt
\font\ninerm=cmr9

\font\eightrm=cmr8

\def\qed{\hbox{${\vcenter{\vbox{			
   \hrule height 0.4pt\hbox{\vrule width 0.4pt height 6pt
   \kern5pt\vrule width 0.4pt}\hrule height 0.4pt}}}$}}

\begin{document}
\begin{flushright}
 UR-1613 \\
 ER/40685/951
\end{flushright}

\normalsize\textlineskip
\thispagestyle{empty}
\setcounter{page}{1}


\vspace*{0.88truein}

\fpage{1}
\centerline{\bf TOP MASS MEASUREMENT}
\vspace*{0.035truein}
\centerline{\bf AND BOTTOM FRAGMENTATION AT THE LHC
\footnote{Talk given at DPF 2000, Meeting of the Division of Particle and 
Fields of the American Physical Society, Columbus, Ohio, 
U.\ S.\ A., 9-12 August 2000.}}
\vspace*{0.37truein}
\centerline{\footnotesize G. CORCELLA}
\vspace*{0.015truein}
\centerline{\footnotesize\it Department of Physics and Astronomy, University
of Rochester}
\baselineskip=10pt
\centerline{\footnotesize\it Rochester, NY 14627,
U.S.A.}

\vspace*{0.21truein}
\abstracts{
We show some recent HERWIG results related to the top quark mass
reconstruction at the LHC and discuss possible improvements
for studies of the bottom quark fragmentation function in the top decay.}{}{}

\textlineskip			
\vspace*{12pt}			
\noindent
The large amount of $t\bar t$ pairs which are expected at the future 
experiments at the LHC$^1$ will allow precise measurements of the 
top-quark properties. 
For the sake of an improved measurement of the top mass 
$m_t$, a new approach has been recently suggested$^2$, which
consists of using the invariant-mass distributions 
of $J/\psi$ + $\ell$
pairs, where the $J/\psi$ is produced by the decay of a $b$-flavoured hadron 
and $\ell$ is the lepton coming from the $W$ produced in the top decay
$t\to bW$.
Even though the branching ratio for the process 
$B\to J/\psi$ is small, it has been shown
that it provides a clean signal at the LHC 
and about $10^3$ well-reconstructed final
states are expected in one year of high luminosity.
The $m_{J/\psi\ell}$ distributions can then be compared with a template of 
shapes parametrized by the top mass and $m_t$ can be fitted.
The expected experimental uncertainty, which is dominated by the statistical
error, has been estimated to be $\Delta m_t\simeq 1$~GeV.
This is a method which crucially relies on the Monte Carlo description of
top production and decay and on the model which is used to
simulate the $b$-quark hadronization.

Standard Monte Carlo algorithms$^{3,4}$ simulate 
multiparton radiation in the soft or collinear approximation and leave
regions of the phase space completely empty (`dead zones').
In particular, HERWIG simulates the top decay in the top rest frame.
We do not have any soft-gluon radiation from the top quark, 
while the $b$ quark is allowed to radiate gluons in the `forward' hemisphere
$0< \theta_g< \pi/2$, with $\theta_g$ being the emission angle of the
soft gluon with respect to the $b$ direction$^5$.
The following parton cascade satisfies the prescription of angular
ordering$^6$.
The total energy loss due to gluon radiation 
is roughly correct, as
the collinear singularity $\theta_g\simeq 0$ 
is correctly taken into account; nevertheless problems are expected
in the angular distributions, since the $W$ `backward' hemisphere 
$\pi/2 < \theta_g < \pi$ is a dead zone for the shower.

The new version HERWIG 6.1$^7$ allows gluon radiation in the backward
hemisphere via the implementation of matrix-element corrections$^8$: 
the dead zone is populated according to the ${\cal O}(\alpha_S)$ matrix element
(`hard correction') and the parton cascade in the already-populated region
is corrected
by the use of the exact amplitude any time an emission is the `hardest
so far' (`soft correction')$^9$.

We wish to use HERWIG 6.1 to study invariant-mass distributions  
which are relevant to extract $m_t$ at the LHC$^{10,11}$.
For simplicity, we study the spectra $m_{B\ell}$, which
must be convoluted with the $B\to J/\psi$ decay spectra
to obtain the $m_{J/\psi\ell}$ distributions. 
In Fig.~1 we show the $m_{B\ell}$ distributions at the LHC according to HERWIG
before (6.0) and after (6.1) matrix-element corrections to top decays,
for $m_t=175$~GeV;
we observe a shift towards lower values of $m_{B\ell}$ after the inclusion
of hard and large-angle gluon radiation. In Table~1 we show results 
for the average values  $\langle m_{B\ell}\rangle$
for different values of $m_t$ 
and observe that the shift induced by matrix-element corrections is of
about 800-900 MeV.
If we parametrize the relation between $\langle m_{B\ell}\rangle$ and
$m_t$ according to a straight line, we find, by means of the least-square 
method: 
\begin{eqnarray} 6.1\ :\;
\langle m_{B\ell}\rangle&=&0.568\ m_t-\ 6.004\ {\mathrm{GeV}}\ ,\
\epsilon({\mathrm{GeV}})=0.057\ ;\\
6.0\ :\;
\langle m_{B\ell}\rangle&=&0.559\ m_t-\ 3.499\ {\mathrm{GeV}}\ ,\
\epsilon({\mathrm{GeV}})=0.052\  ,
\end{eqnarray}
$\epsilon$ being the mean square deviation in the fit.

Inverting the above relations, we find that the values of $m_t$ extracted
using the two versions HERWIG 6.0 and 6.1 differ by about 1.5 GeV, a
value which is larger than the expected uncertainty.
This implies that, for an accurate reconstruction of the top mass, the
corrections to the top decay must be applied.
If we set a cut  $m_{B\ell}>50$~GeV
to reduce the effect of possible backgrounds on the low-mass tails, the 
impact of matrix-element corrections to top decays is reduced to 1 GeV, but
it is still competitive with respect to the expected uncertainty$^{11}$.
\begin{figure}[htbp]
\centerline{
\epsfig{file=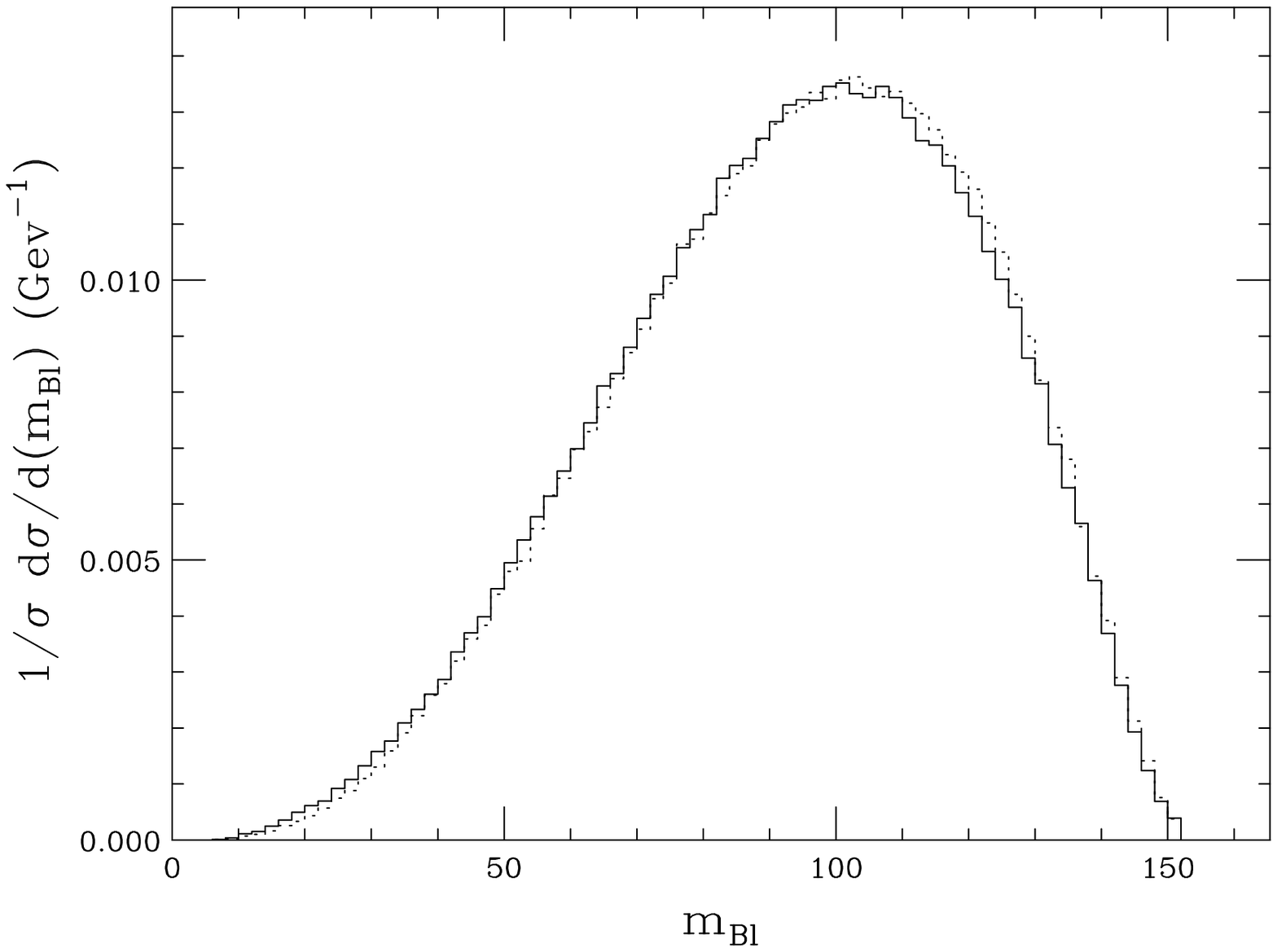,height=2.5in,width=3.5in,clip=}}
\fcaption{$m_{B\ell}$ distributions at the LHC according 
to HERWIG 6.1 (solid histogram) and 6.0 (dotted).}
\end{figure}
\begin{table}[htbp]
\tcaption{Results for the invariant mass $m_{B\ell}$ at the LHC for 
different values of $m_t$.}
\centerline{\footnotesize\smalllineskip
\begin{tabular}{l c c c c c}\\
\hline
 $m_t$&$\langle m_{B\ell}\rangle^{6.1}$&$\sigma(6.1)$&
$\langle m_{B\ell}\rangle^{6.0}$&
$\sigma(6.0)$&$\langle m_{Bl}\rangle^{6.0}-\langle  
m_{Bl}\rangle^{6.1}$\\
\hline
171 GeV&91.13 GeV&26.57 GeV&92.02 GeV&26.24 GeV&$(0.891\pm 0.038)$  
GeV\\\hline
173 GeV&92.42 GeV&26.90 GeV&93.26 GeV&26.59 GeV&$(0.844\pm 0.038)$  
GeV\\\hline
175 GeV&93.54 GeV&27.29 GeV&94.38 GeV&27.02 GeV&$(0.843\pm 0.039)$  
GeV\\\hline
177 GeV&94.61 GeV&27.66 GeV&95.46 GeV&27.33 GeV&$(0.855\pm 0.039)$  
GeV\\\hline
179 GeV&95.72 GeV&28.04 GeV&96.51 GeV&27.67 GeV&$(0.792\pm 0.040)$  
GeV\\\hline
\end{tabular}}
\end{table}

The next step of the presented analysis is a more detailed
study of the $b$ fragmentation in the top decay$^{12}$.
Another possible approach to obtain the $m_{B\ell}$ distributions 
is to perform
an exact next-to-leading order
calculation of the decay rate $t\to bW(g)$ and convolute the result
with the $b$ fragmentation function, taken from some
LEP or SLD data$^{13}$. 
It will be very interesting to compare the HERWIG results to the ones
of this calculation. 
In order for such a comparison to be
trustworthy, we have to tune the parameters of the HERWIG cluster model
to fit the $e^+e^-$ data. We have a recent tuning by the OPAL 
collaboration$^{14}$, after which the agreement of HERWIG to the LEP data
is considerably improved. However, even after such a tuning, problems 
have been found in the comparison to the SLD data$^{15}$, which 
was not taken into account as an input for the OPAL tuning.
The tuning of the HERWIG hadronization model 
using  both LEP and SLD data is in progress$^{12}$. 

In summary, we reviewed some HERWIG results which are relevant to the top
mass determination at the LHC using final states with leptons and $J/\psi$
and found that the effect of matrix-element corrections to top 
decays is $\Delta m_t\simeq 1.5$~GeV.
We also reported on further studies for a deeper understanding of the 
$b$ fragmentation in the top decay, using data from the 
$e^+e^-$ machines.

\nonumsection{Acknowledgements}
\noindent
The presented results have been obtained in collaboration with M.L. Mangano and
M.H. Seymour. We acknowledge A. Kharchilava and R. Hemingway for useful
discussions and correspondance.
This work was supported by grant number DE-FG02-91ER40685 from the U.S.
Department of Energy.

\nonumsection{References}

\end{document}